\documentclass[prl,twocolumn,showpacs,superscriptaddress,floatfix,amsmath,amssymb]{revtex4}

\usepackage{graphicx}

\begin{document}
\title{\bf From local to macroscopic coherence in systems with composite
quasi-particles}
\date{\today}
\author{J.Ranninger}
\affiliation{Centre de Recherches sur les Tr\`es Basses
Temp\'eratures, Laboratoire Associ\'e \`a l'Universit\'e Joseph
Fourier, Centre National de la Recherche Scientifique, BP 166,
38042, Grenoble C\'edex 9, France}
\author{A.Romano}
\affiliation{Dipartimento di Fisica "E.R. Caianiello",
Universit\`a di Salerno, I-84081 Baronissi (Salerno), Italy -
Unit\`a I.N.F.M. di Salerno}

\begin{abstract}
Strongly interacting systems are characterized by heavily dressed
entities with internal degrees of freedom, which, on a local
level, can be described in terms of coherent quantum states. We
examine the modification of these {\it local} coherent quantum
states when such entities condense into a {\it macroscopic}
coherent quantum state, such as superfluidity. As an example, we
consider a system of electrons coupled to local lattice
deformations. Significant changes in the phonon clouds surrounding
the charge carriers occur when the system develops into a
spatially phase-locked state. The question of localized
self-trapped charge carriers (bipolarons) in the normal state
becoming delocalized upon entering the superconducting phase is
discussed in terms of squeezing of the local coherent phonon
states. Suggestions for experimental verifications of these
features associated with the lattice dynamics are made.

\end{abstract}


\maketitle

\section{INTRODUCTION}
Particles which locally strongly interact with their environment
polarize this latter and finish up in self-trapped states. On a
local level, the environment surrounding the particles can be
described in terms of a coherent superposition (in form of Glauber
states) of the elementary bosonic excitations of the bare
environment. As a result, new composite entities form, given by
the charge carriers and their surrounding clouds of bosonic
excitations of the environment. It is at present a matter of
dispute whether in the normal state such composite entities can
exist as itinerant Bloch-like states or are purely diffusive. The
retarded interaction between the charge carriers and the bosonic
excitations of the  environment favors the picture of diffusive
motion caused by  a dephasing of the dynamics of the two constituents.

The question we want to address here is whether this process of
dephasing can be blocked, leading to a delocalization when such
self-trapped composite entities condense into  a {\it macroscopic}
coherent quantum state. We  investigate such a possibility in
terms of a  model of very general form, given by bosonic charge
carriers strongly coupled to local bosonic excitations of the
environment. The bosonic charge carriers can, amongst others,
represent electron pairs induced by strong electron-phonon or
electron-spin fluctuation interaction, electron pairs in double
valence fluctuation systems (stabilized by their ligand
environment), Frenkel excitons (stabilized by their interaction
with the surrounding lattice deformations) and, in a more remote
sense, He$^4$ atoms in a bath of He$^3$ with whose excitations
they interact. Such bosonic entities can in principle condense
into a superfluid state. Our interest here is to show how in  such
a {\it macroscopic} coherent quantum state  the internal structure
of the {\it local} coherent quantum state of the bosonic
excitations of the environment is modified.

We shall investigate this problem on the basis of a model of
itinerant electrons, coupled to some purely local vibrational
modes. For moderately strong and strong coupling, it results in
self-trapped composite entities, given by on-site electron pairs,
surrounded by local lattice deformations which are described by
clouds of phonons in local coherent quantum states. In the
intermediate coupling limit these localized entities can be
considered as two-particle resonant states of the underlying system of
itinerant electrons. They form out of pairs of such uncorrelated
itinerant electrons and ultimately decay into them.

\section{THE BOSON-FERMION MODEL}

A model Hamiltonian which captures such physics is given by the
boson-fermion model (BFM)
\begin{eqnarray}
&&H_{\rm BFM}  =  (D-\mu)\sum_{i,\sigma}n_{i\sigma}
-t\sum_{\langle i\neq j\rangle,\sigma}c^+_{i\sigma}c_{j\sigma}
\quad \nonumber \\
&&+  (\Delta_B-2\mu) \sum_i \left( \rho_i^z + \frac{1}{2} \right)
+v\sum_i [\rho^+_ic_{i\downarrow}c_{i\uparrow}
+\rho_i^- c^+_{i\uparrow}c^+_{i\downarrow}] \nonumber \\
&&-  \hbar \omega_0 \alpha \sum_i \left( \rho_i^z + \frac{1}{2}
\right) (a_i+a_i^{+}) +\hbar \omega_0 \sum_i \left(a^{+}_i a_i
+\frac{1}{2}\right) \, . \nonumber \\
\end{eqnarray}

Here $\rho_i^{\pm}$ denote the creation and annihilation operators
of hard-core bosons (characterizing the self-trapped localized
electron-pairs) on some effective sites $i$, which should be
understood as of being made up of small clusters of atoms rather
than single atomic sites. Such hard-core bosons have
spin-$\frac{1}{2}$ commutation relations $\left[ \rho_i^-,\rho_i^+
\right]_-=2 \rho_i^z$ and $\left[ \rho_i^-,\rho_j^+ \right]_+=
\delta_{ij}$. $c_{i\sigma}^{(+)}$ denote the fermionic operators
referring to annihilation (creation) of itinerant electrons with
spin $\sigma$, $n_{i\sigma}=c^+_{i\sigma}c_{i\sigma}$ being
the number operator for such fermions. $a_i^{(+)}$ denote
annihilation (creation) operators of the excitations of the
environment, related to the local lattice displacements
$X_i=(a_i+a_i^+)/\sqrt{2M\omega_0/\hbar}$ where $M$ is some atomic
mass characterizing the effective sites, and $\omega_0$ is the
frequency of the local deformations of the lattice. The bare
hopping integral for the itinerant electrons is given by $t$, the
bare electron half bandwidth by $D=zt$ (which will serve as energy
unit), $z$ denoting the coordination number. $\alpha$ denotes
the coupling of the hard-core bosons to the deformations of the
surrounding medium and $v$ the exchange coupling between the bosons
and the pairs of itinerant fermions. The bare boson energy level is
denoted by $\Delta_B$. The chemical potential $\mu$ is common to
fermions and bosons such as to guarantee overall charge conservation.

The model defined by the Hamiltonian (1) was introduced shortly
after the proposition of the concept of {\it bipolaronic
superconductivity} \cite{Alexandrov-81} (appropriate to the
extreme strong coupling anti-adiabatic regime) in an attempt to
extend this concept to the regime of intermediate electron-lattice
deformation coupling. This model is based on a conjecture, rather
than on a derivation from an underlying basic electron-phonon
Hamiltonian, such as for instance the Holstein
model\cite{Holstein-59}.

We shall in the following develop the ideas which had initially
led us to this conjecture. Let us start from a simple Holstein
model with purely local Einstein modes for the lattice vibrations. The
corresponding Hamiltonian is given by
\begin{eqnarray}
H&=&(D-\mu)\sum_{i\sigma}n_{i\sigma}-
t\sum_{i \neq j,\sigma}(c^{+}_{i\sigma}c^{\phantom{+}}_{j\sigma}+
h.c.) +
U\sum_i n_{i\uparrow}n_{i\downarrow} \nonumber \\
&&-\bar \alpha \hbar \omega_0\sum_{i,\sigma}
n_{i\sigma}(a_i+a^{+}_i) + \sum_i \hbar \omega_0 \left(a^{+}_ia_i
+\frac{1}{2}\right)  \; . \label{H}
\end{eqnarray}
The meaning of the various operators and coupling constants is the
same as in the Hamiltonian (1), except for the value of $\bar
\alpha$ which is equal to $\frac{1}{2} \alpha$.

The electron-phonon interaction  with coupling strength $\bar
\alpha$ is taken to be local, in view of the interaction with
primarily local phonon modes  which favours small polaron
formation. $U$ denotes some effective Coulomb repulsion - not to
be confused with a Hubbard-type on-site interaction - on the
effective sites, which, we stress again, ought to be considered as
being composed of small molecular clusters such as a diatomic
molecular unit or similar. Electrons described by this Hamiltonian
can gain energy either by becoming itinerant and remaining
essentially uncoupled to the lattice or, alternatively, getting
self-trapped. In the first case the energy gain is described by
the intrinsic band dispersion
\begin{equation}
\varepsilon_{\bf k}= D(1-\gamma_{\bf k}) \qquad \gamma_{\bf
k}=\frac{1}{z}\sum_{\bf \delta} e^{i{\bf k\delta}} \; ,
\end{equation}
${\bf \delta}$ denoting lattice vectors linking nearest-neighbor
sites. For self-trapped polaronic states, the local part of the
Hamiltonian (\ref{H}) can be transformed by a shift in the phonon
variables $a^{(+)}_i\rightarrow a^{(+)}_i+ \bar\alpha
(n_{i\uparrow} +n_{i\downarrow})$ upon which it acquires the form
\begin{eqnarray}
H_{loc}&=&(D-\mu-\varepsilon_p)\sum_{i\sigma}n_{i\sigma}-
(2\varepsilon_p- U)\sum_i n_{i\uparrow}n_{i\downarrow} \nonumber \\
&&+\sum_i \hbar \omega_0 \left(a^{+}_ia_i +{1\over 2}\right) \; ,
\label{Hloc}
\end{eqnarray}
with the polaron binding energy
$\varepsilon_p = \bar\alpha^2 \hbar \omega_0$
denoting the gain in energy by the corresponding polaron
level shift. The effective on-site interaction given by
$(2\varepsilon_p- U)$ controls the possible formation of
bipolaronic states
\begin{equation}
\rho^{+}_i \, X_i^{+}\equiv c^+_{i\uparrow}\,c^+_{i\downarrow}\,
e^{-2\bar\alpha(a_i - a^+_i)}
\end{equation}
when $2\varepsilon_p- U > 0$, with the single polaron states lying
above such bipolaron states. For $2\varepsilon_p- U < 0$ we have
the inverse situation with the bipolaronic level lying above that
of the single polarons. What we are interested here is the first
case.

Let us begin by considering the situation of sufficiently strong
interaction such that the bipolaronic level lies below the bottom
of the unrenormalized electron band, i.e., for $2\varepsilon_p-
\frac{1}{2}U > D$, as illustrated in  Fig.\,1. In that case
bipolarons can acquire itinerancy by decaying into virtual states
of pairs of itinerant electrons and reassembling subsequently on
some neighbouring site. Within the usual Lang-Firsov
approximation, neglecting  phonon creation as well as annihilation
processes during the charge transfer from one site to the other,
the corresponding effective bipolaron hopping integral is then
given by $t^{**}\simeq t^2e^{-2\bar\alpha^2}/(2\varepsilon_p -
\frac{1}{2}U-D)$\cite{Alexandrov-81}. Within this approximation
one obtains a system of itinerant bipolarons on a lattice which,
at low temperatures, can condense into a superfluid state, giving
rise to what has been termed {\it bipolaronic superconductivity}.
This theoretical proposition has ever since remained a matter of
dispute, questioning if such a normal-state itinerant Bose liquid
can be achieved in real materials. The main arguments against it
is that the stability of local bipolarons requires a relatively
high value of $\bar\alpha^2 \geq (D +\frac{1}{2}U)/2\hbar
\omega_0$ for realistic values of $D$, $U$ and $\omega_0$ and
hence leads to exceedingly small values of the bipolaron
bandwidth. A further hampering factor for the realization of such
bipolaronic itinerant band states is related to the retardation
effects between the motion of the charge carriers and the lattice
deformations which surround them. This gives rise to dephasing
between the two, as shown by exact diagonalization
approaches\cite{deMello-97} and DMRG studies\cite{Zhang-99},
indicating that incoherent processes in the optical conductivity
are dominant. All these results suggest that coherent bipolaron
motion is most unlikely to occur in real systems and it was for
that reason that the Boson-Fermion model was initially
proposed\cite{Ranninger-85}. The aim was to capture the physics
outside the regime of the extreme strong coupling antiadiabatic
limit and to describe the intermediate coupling case, where the
bipolaron level lies inside the band of the itinerant electrons
(see Fig.\,2) and where the superconducting state is controlled by
the phase correlations of these resonant bosonic bipolaron states.

Quite  generally, this model can be considered as a paradigm for
cross-over situations. For the specific case we are interested
here, it describes a cross-over between a BCS-type superconductor
of Cooper pairs in the weak coupling regime and the hypothetical,
yet to be experimentally verified, {\it bipolaronic
superconductivity} in the strong coupling regime. In a totally
different physical situation, namely that of a strongly correlated
electron system such as described by the Hubbard model, it has
been recently shown that the intermediate coupling case can be
mapped onto  such a Boson-Fermion scenario\cite{Auerbach-01}. Such
an effective Boson-Fermion Hamiltonian was also derived recently
for the exchange interaction between spinon singlets of the RVB
electron-pairs and holons\cite{Kochetov-02}. The question to what
extent such a BFM can emerge from a quite general class of
interacting fermion systems, has been addressed from the point of
view of  a bosonization procedure of an intrinsically fermionic
system\cite{Lee-94}. The presently much discussed molecular
Bose-Einstein condensate, involving entangled atoms in squeezed
states has also recently been analyzed\cite{Yurovski-02} in the
framework of such a BFM.

A derivation of the BFM for moderately strongly coupled
electron-phonon systems, starting from the Holstein model, is so
far not available. In order to provide the physically intuitive
picture which initially had led us to propose this model, let us
consider the case where the local bipolaron states are stable
relative to single polarons. This eventuality can only be decided
on the basis of direct experimental evidence for such local
bipolaronic entities. The classical cases which initially had led
to consider such a situation are systems like amorphous
semiconductors\cite{Anderson-75}, Ti$_4$O$_7$\cite{Schlenker-76},
Na$_x$V$_2$O$_5$\cite{Chakraverty-78}, WO$_{3-x}$\cite{Salje-80}
and double valence fluctuation systems\cite{Chernick-82} like
PbTe\,(Tl), just to name a few of them, and all of which represent
such bipolarons as {\it exclusively} localized self-trapped
states.

When these systems are doped, the electrons responsible for pairing
are at the same time responsible for the - albeit diffusive -
transport. It is hence tempting to hypothesize that such
bipolaronic states can occur as resonant states inside a band of
itinerant electrons. In more complex systems, such as ternary and
quaternary compounds, different electrons come into play. Some of
them are prone to form bipolarons and others will remain
itinerant, but being coupled together via some
hybridization term. In systems like that, such a Boson-Fermion
scenario imposes itself more directly. It is the eventuality of
bipolaronic resonant states rather than localized states, which
was at the origin of our proposal of the Boson-Fermion
scenario\cite{Ranninger-85}.

\begin{figure}
\includegraphics[width=6cm]{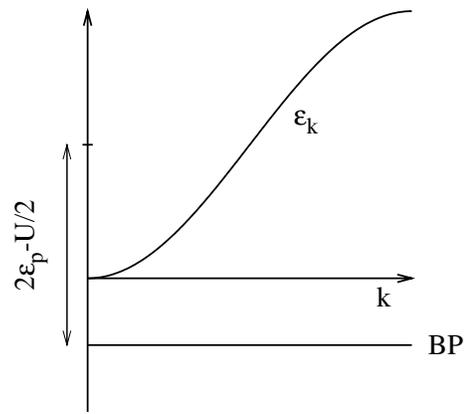} \caption{Schematic
plot of the case of the bipolaron level (BP) falling below the
band of itinerant electrons.} \label{band1}
\end{figure}
\begin{figure}
\includegraphics[width=6cm]{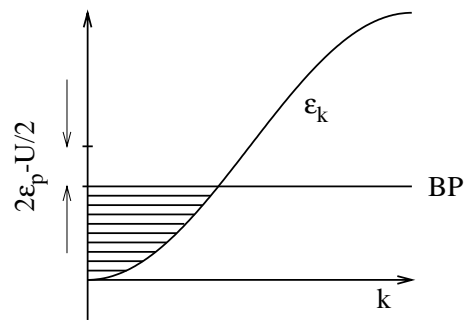} \caption{Schematic
plot of the case of the bipolaron level (BP) falling inside the
band of itinerant electrons.} \label{band2}
\end{figure}

The level scheme for the local bipolaronic states inside the band
of the bare itinerant electrons for this situation is depicted in
Fig.\,2. Filling up such a system gradually with electrons, it is
evident from Fig.\,2 that, as long as the Fermi level remains
below the level of the local bipolaronic states, the charge
carriers will be essentially bare itinerant electrons. But upon
further increasing the concentration of charge carriers, the
itinerant electrons will eventually overlap in energy with the
intrinsically localized bipolaronic level.

The major supposition leading to this BFM is now to assume that
this overlap can be simulated by some effective on-site
hybridization describing a charge exchange between the  bipolarons
and pairs of itinerant electrons in the form of
\begin{equation}
v\sum_i [\rho^+_iX_i^+c_{i\downarrow}c_{i\uparrow} +\rho_i^-
X_i^-c^+_{i\uparrow}c^+_{i\downarrow}] \; ,
\end{equation}
with the itinerant electrons and the localized bipolarons,
Eq.\,(5), treated as commuting fields. Such a supposition is supported by
the fact that the spectrum of electrons, moderately strongly
coupled to local modes, is known to be given by the bare spectrum
for all ${\bf k}$-vectors, except where their energy crosses the
level of localized selftrapped electrons. In this regime of
energies the electron spectrum is strongly modified, as shown by
exact diagonalization studies\cite{Fehske-97,Stephan-97} as well
as variational methods\cite{Romero-99}, with the spectrum becoming
rather flat and  approaching the local level of the self-trapped
polaronic states. These are features which are characteristic of
systems where itinerant charge carriers are hybridized with
localized charge carriers, and this was the center of our initial
assumption of an effective hybridization between localized
bipolarons and pairs of itinerant electrons, as given by Eq.\,(6).

On a local level the dynamics of a polaronic system, involving
two-particle resonant states, is controlled by the atomic limit of
such a system, which we thus propose to describe by the following
Hamiltonian
\begin{widetext}
\begin{eqnarray}
H_{at} & = &(-4\varepsilon_p+U-2\mu) \sum_i
\left(\rho_i^z+\frac{1}{2} \right)+ v\sum_i
\left[\rho^+_iX_i^+c_{i\downarrow}c_{i\uparrow} + \rho_i^-
X_i^-c^+_{i\uparrow}c^+_{i\downarrow}\right] +\sum_i \hbar
\omega_0 \left(a^{+}_ia_i +\frac{1}{2}\right)
\\
&\equiv &(U-2\mu) \sum_i \left(\rho_i^z+\frac{1}{2}\right)+
v\sum_i \left[\rho^+_ic_{i\downarrow}c_{i\uparrow} +\rho_i^-
c^+_{i\uparrow}c^+_{i\downarrow}\right]- \hbar \omega_0 \alpha
\sum_i \left(\rho_i^z + \frac{1}{2}\right)
\left(a_i+a_i^{+}\right)+\sum_i \hbar \omega_0 \left(a^{+}_ia_i +
\frac{1}{2}\right) , \nonumber
\end{eqnarray}
\end{widetext}
with a bare bipolaron level given by $\Delta_B=U$.

The equivalent formulation of this Hamiltonian [the second part of
Eq.\,(7)] is obtained after employing an inverse Lang-Firsov-type
transformation $H_{at} \rightarrow U^+ H_{at} U$ with
$U=e^{\alpha(a_i - a^+_i)(\rho^z_i+\frac{1}{2})}$. Adding to this
local Hamiltonian the term describing the kinetic energy of the
itinerant electrons, leads immediately to the BFM Hamiltonian
given by Eq.\,(1).

\section{COMPETITION BETWEEN LOCAL AND  GLOBAL COHERENCE}
\subsection{THE PSEUDOGAP PHASE}

The basic physics contained in  this BFM model originates from a
competition of the local electron pairing, acting as a prerequisite
for superconductivity, with the delocalizing effect of the itinerant
electrons which favors non-local Cooper pairing as the temperature
is lowered and the superconducting phase is approached. The
change-over from local pairing to Cooper pairing is not only
manifest in the electronic properties of such systems but is
expected to be accompanied by corresponding lattice deformations,
the study of which is the issue of the present paper.

We shall examine this feature for the two following extreme
limits:

i) the normal-state phase, characterized by the existence of a
pseudogap in the density of states (DOS) of the itinerant
electrons, which will be treated here in an approximate fashion
within the atomic limit of the BFM,

ii) the superconducting phase, which will be treated on a
mean-field level, taking into account the phase locking of the
bound electron pairs.

Throughout this study we choose as a representative example a set of
parameters presenting a situation close to the fully symmetric case. This
implies an effective bipolaron level (corresponding to the energy level
for the bound pairs) given by $\Delta_B - 4 \varepsilon_p \simeq 0$,
resulting in a final state  bipolarons level lying close to the
center of the band of the itinerant electrons, after the boson-phonon
coupling has been switched on. We moreover want to consider the case
$n_{tot}=2n_F + n_B = 2$, giving a concentration of electrons
$n_F \simeq 0.5$ and a concentration of bound pairs $n_B \simeq 0.5$
and choose $v=0.25$, $\omega_0=0.1$ and $\alpha=2$.

Let us start from the purely local physics, described by the
atomic limit ($t=0$) of the BFM corresponding to the Hamiltonian
$H_{at}$, Eq.\,(7). This limit has been studied by us
previously\cite{Ranninger-98a} in connection with the pseudogap
phenomenon and the incoherent background in the electron spectral
function, which, in such a scenario, arise from phonon shake-off
processes.  For the choice of parameters adopted here, the physics
controlling the change-over from the normal into the
superconducting phase can essentially be represented by the one-,
two- and three-particle eigenstates of $H_{at}$
\begin{eqnarray}
|1,l\}^{at}_{\sigma} &=& c^+_{\sigma}|0\rangle|0) |l>\\
|2,l\}^{at} &=& \sum_n \left (u^{at}_{l n}\,c^+_{\uparrow}
c^+_{\downarrow} + v^{at}_{ln}\,\rho^+ \right )|0\rangle|0) |n> \\
|3,l\}^{at}_{\sigma} &=& \sum_n \ c^+_{\sigma}\rho^+|0\rangle|0)
{(a^+-\alpha)^l \over \sqrt {l!}}e^{-\frac{1}{2}\alpha^2}{\alpha^n
\over \sqrt {n!}} |n>
\end{eqnarray}
with $|i,j\}^{at}$ denoting the $i$-particle $j$-th eigenstate.
 The opening of the pseudogap at a certain temperature
$T^*$ shows up in the rapid spectral weight increase, with
decreasing temperature below $T^*$, of the scattering amplitude
involving transitions from the lowest energy two-particle bonding
state $|2,0\}^{at}$ to the lowest energy single-particle states
$|1,0\}^{at}_{\sigma}$ and from the lowest energy three-particle
states $|3,0\}^{at}_{\sigma}$ to $|2,0\}^{at}$. These
contributions to the DOS of the itinerant electrons are manifest
by the wings (see Fig.\,3) for energies below and above that of
the single particle non-bonding contributions, which have spectral
weight around the chemical potential (equal to $\omega=0$)
and which arise from transitions not involving any of the
two-particle states. Decreasing the temperature below $T^*$, the
spectral weight at $\omega=0$ decreases rapidly and transfers
to the wings of the DOS, thus resulting in the opening of a
pseudogap. Taking into account the dispersion of the itinerant
electrons in an approximate fashion,  such as within a Coherent
Potential Approximation (CPA) approach\cite{Domanski-01a}, the
main features seen in Fig.\,3 are maintained, simply leading to a
smearing of the DOS.
\begin{figure}
\includegraphics[width=8cm]{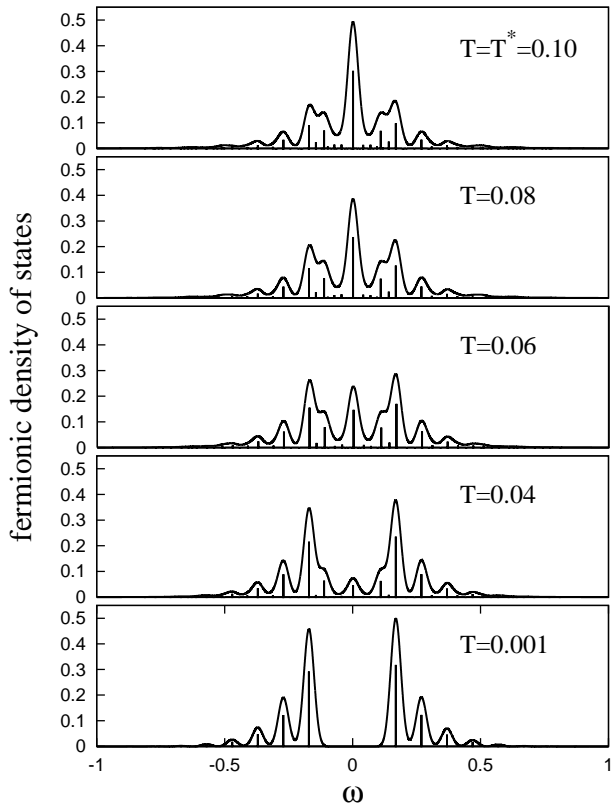} \caption{Evolution
of the fermionic DOS with temperature, showing the opening of the
pseudogap at the chemical potential ($\omega=0$), as the temperature
decreases below $T^*$.}
\label{DOS}
\end{figure}
The two-particle eigenstates of $H_{at}$, Eq.\,(9), are obtained by
expanding the local oscillator states in terms of the eigenstates
$|n>_i$ of the undeformed harmonic oscillator at a given site $i$.
$|0\rangle$ and $|0)$ denote respectively the vacuum states of the
itinerant electrons and of the bound electron-pairs on such a
site. The coefficients $u^{at}_{l n}$ and $v^{at}_{l
n}$ are determined by exact diagonalization of this atomic
limit problem within a restricted Hilbert space of those excited
oscillator states. It is generally sufficient to take into account
up to $50$ such states in order to get converging results.

The qualitative changes in the lattice deformations expected to
occur when going from the pseudogap phase into the superconducting
phase are essentially described by the lowest energy one-, two-
and three-particle states, which contain the information on the
relative weight of the bound electron-pairs ($v_{ln}^{at}$) and of
the induced pairing in the subsystem of the itinerant electrons
($u_{ln}^{at}$). The first ones are intrinsically coupled to the
local lattice deformations while the second ones acquire such
local lattice deformations due to the charge exchange mechanism
acting between the two. As we shall see below, the efficiency of
this transfer of polaronic features is noticeably different in the
normal and in the superconducting phase. It shows up in the
composition $P(n)$ of the coherent phonon states describing the
local phonon clouds surrounding the charge carriers on a given
site $i$, which in the normal state (atomic limit) is given by
\begin{equation}
P_{at}(n) = {1 \over Z}\sum_{m,l}  exp(-\beta E^{at}_{ml}) \;
^{at}\{m,l|a^+a |m,l\}^{at} \; ,
\end{equation}
$Z$ denoting the partition function and $E^{at}_{ml}$ the
eigenvalues of $H_{at}$ associated with the eigenstates
$|m,l\}^{at}$. Upon decreasing the temperature, $P_{at}(n)$ shows
a redistribution of weight from high to small values of the phonon
number $n$ [see Fig.\,4(a)].
\begin{figure}
\includegraphics[width=8cm]{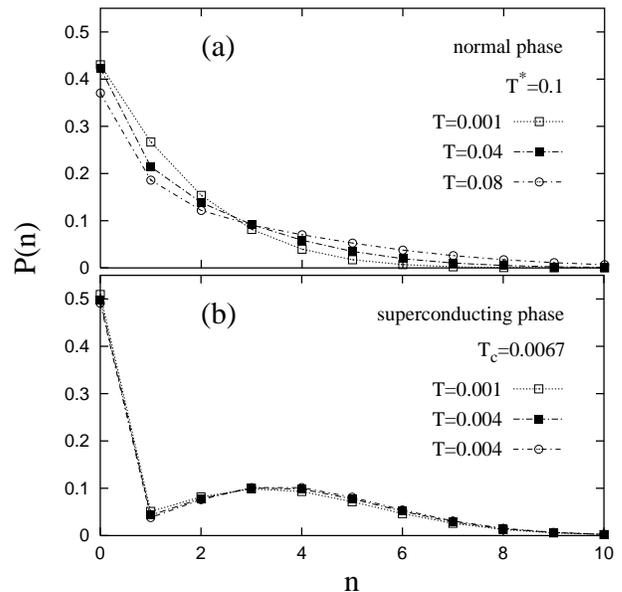} \caption{Comparison
of $P(n)$ in the atomic limit and in the superconducting state for
different temperatures $T$.} \label{P(N)0}
\end{figure}
In the temperature regime of interest here, i.e., from slightly
above $T^*$ all the way down to $T=0$, the contributions to $P(n)$
come essentially from the lowest energy one-, two- and
three-particle states, Eqs.\,(8-10). As illustrated in Fig.\,5,
these various contributions have noticeably different features as
far as the phonon distributions are concerned. The one-particle
states are totally decoupled from the lattice and hence have a
contribution only for $n=0$, while the three-particle contribution
are maximally coupled to the lattice showing a broad peak in this
distribution which corresponds to that of a shifted oscillator
ground state. The two-particle bonding state $|2,0\}^{at}$ has
features of a displaced oscillator which is particularly
noticeable at low temperatures where it is the main contribution,
together with a much reduced contribution from the two-particle
antibonding state $|2,1\}^{at}$. That latter contribution to
$P(n)$, considered without thermal weighting factor, clearly shows
strong contributions for low values of $n$, which is indicative of
weak correlations between the electronic degrees of freedom and
those of the lattice.
\begin{figure}
\includegraphics[width=7cm]{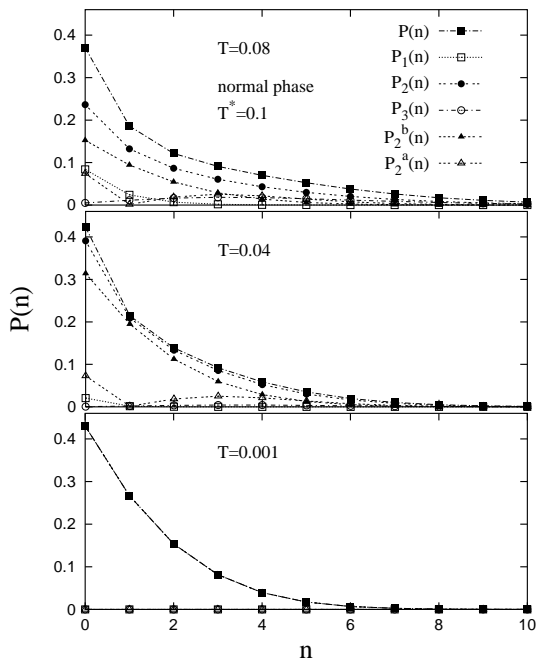}
\caption{Decomposition, for different temperatures, of $P(n)$ in
the normal state (full squares) into  contributions coming from
the one-, two- and three-particle states ($P_1(n)$, $P_{2}(n)$ and
$P_{3}(n)$, respectively). $P_{2}^b(n)$ and $P_{2}^a(n)$ represent
the contributions coming respectively from the lowest energy
two-particle state (bonding) and from the first excited
two-particle state (antibonding).} \label{P(N)1}
\end{figure}
The onset of the local coherence between bound pairs and of the
induced pairing in the itinerant subsystem can be tracked in the
pair distribution function (PDF) $g(X)$, given in the normal state
(atomic limit) by
\begin{equation}
g_{at}(X) = {1 \over Z} \sum_{m,l} e^{-\beta E^{at}_{ml}} \;
^{at}_i\{m,l| \delta (X - X_i)  | m,l \}^{at} _i \; ,
\end{equation}
which describes the probability of finding a local deformation of
size $X$ and which can be studied by EXAFS and pulsed neutron
scattering experiments\cite{Egami-96}. $g_{at}(X)$ is obtained by
expressing the eigenstates of $H_{at}$ in terms of a real-space
representation involving the harmonic oscillator wavefunctions.
In Fig.\,6(a) we illustrate the temperature evolution of the PDF.
Its decomposition into the contributions arising from
\begin{figure}
\includegraphics[width=7cm]{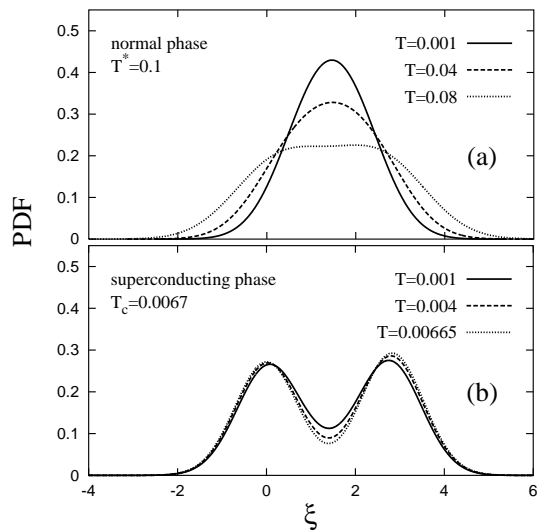} \caption{Comparison
of the PDF as a function of the displacement $X$, measured by the
dimensionless parameter $\xi= X\sqrt{M\omega_0/\hbar}$, in the
atomic limit and in the superconducting phase for different $T$.}
\label{PDF0}
\end{figure}
\begin{figure}
\includegraphics[width=8cm]{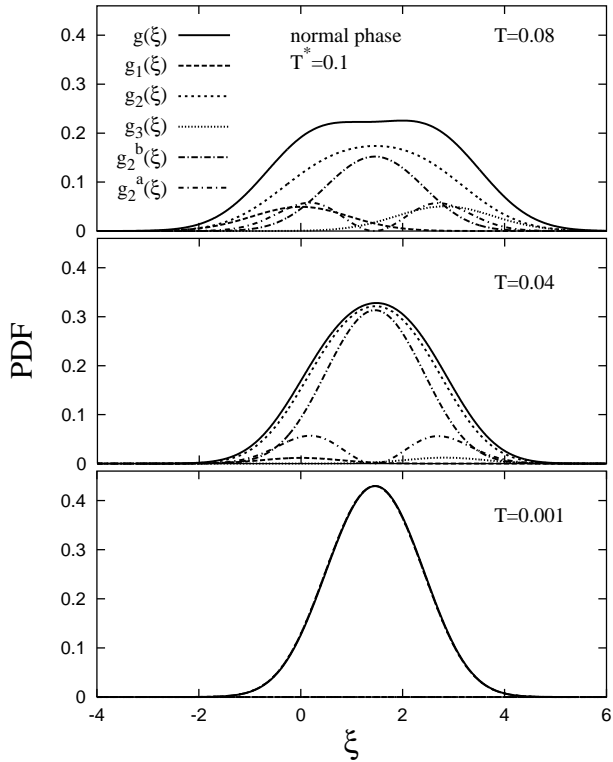}
\caption{Decomposition, for different temperatures, of the PDF in
the normal phase (full lines) into contributions coming from the
one-, two- and three-particle states ($g_1(\xi)$, $g_2(\xi)$ and
$g_3(\xi)$, respectively). $g_{2}^b(\xi)$ and $g_{2}^a(\xi)$
represent the contributions coming respectively from the two
lowest bonding and antibonding two-particle eigenstates of
$H_{at}$.} \label{PDF1}
\end{figure}
\noindent the one-, two- and three- particle states given in
Eqs.\,(8-10) are illustrated in Fig.\,7. The one- and
three-particle states contribute to the PDF in form of  peaks
centered around $X=0$ and around some finite value of $X$,
respectively. This expresses the fact that the one-particle states
(8) are completely decoupled from the lattice, while the
three-particle states (10) contain a bound electron-pair and thus
give a maximal shift in the PDF. The two-particle bonding and
anti-bonding states, Eq.\,(9) with $l=0,1$, respectively, give
contributions to the PDF with two peaks lying very close together,
which is a result of the strong correlation between the bound
electron pairs and the induced pairs in the itinerant electron
system, characterizing the normal-state phase. At high
temperatures, the PDF is thus given by a very broadened smeared
out double-peak structure.  But as the temperature is lowered
below $T^*$, local coherence between the bound electron pairs and
the pairs of the itinerant electrons sets in. As a result, the
opening of the pseudogap in the local density of states is
accompanied by a change-over in the PDF from this very smeared out
double-peak structure to a rather well defined single-peak
structure. The latter is essentially due to the lowest energy
bonding two-particle eigenstate contributing to a PDF with a
maximum at some finite value of $X$. This is a signature of the
strong correlations, which gradually build up as we go below
$T^*$, between the bound pairs and the induced pairing in the
itinerant electron subsystem leading to  roughly equal weight
coming from the coefficients $u_{ln}^{at}$ and $v_{ln}^{at}$.

\subsection{THE SUPERCONDUCTING PHASE}

Let us now  put such single units, as described by $H_{at}$,
together in an infinite lattice, and let them  interact with each
other via the hopping term of the itinerant electrons. When the
system undergoes a transition into a superfluid state of the
charge carriers, the on-site correlations between bound pairs and
pairs of itinerant electrons will be weakened, since the ${\bf
k}$-space pairing of the itinerant electrons enters into
competition with the strong on-site pairing which controls the
physics in the normal state. As a result, upon entering the
superconducting phase, a substantial modification of the coherent
phonon states of the individual units will occur.

We shall,for a moment, make the assumption that the deformation of
the medium surrounding the bosonic charge carriers is not
influenced by boson-fermion charge fluctuations. In that case the
coupling of the bound electron-pairs to the lattice deformations
can be eliminated, along the lines usually adopted in studies of
the polaron problem, leaving us with an effective Hamiltonian
given by the first four terms of the Hamiltonian (1) and with a
renormalized hybridization coupling $ve^{-\alpha^2/2}$. Such a
Hamiltonian has been studied in detail within different schemes,
such as self consistent diagramatic techniques\cite{Ranninger-95},
dynamical mean-field theory\cite{Ranninger-98b} and
renormalization group procedures\cite{Domanski-01b}, and gives
rise to the following physics. As the temperature is lowered below
a certain characteristic temperature $T^*$, pairing of the
itinerant electrons sets in, leading to an emptying out of the
single electron spectrum near the chemical potential and thus to
the opening of a pseudogap. Upon further decreasing the
temperature these electron-pairs become
itinerant\cite{Devillard-00}, eventually condensing into a
macroscopic superfluid ground state of the form
\begin{equation}
\prod_i [u_ie^{-i\phi_i/2} + v_i e^{+i\phi_i/2}\rho_i^+]|0)_i
\end{equation}
($u_i, v_i$ being some model dependent parameters) and exhibiting
soundwave-like collective excitations as a consequence of the
rigidity of the local phases $\phi_i$.

In order to study how such a superfluid state affects the
surrounding lattice deformations, we now investigate this problem
on the basis of a mean-field treatment of the full Hamiltonian
with the coupling of the phonons to the bound pairs explicitly
taken into account. To this purpose, we introduce the following
order parameters for the charge carriers\cite{Ranninger-99}
\begin{equation}
x \; = \; {1 \over N} \sum_i \langle c_{i \uparrow}^+ c_{i
\downarrow}^+ \rangle \, , \quad \rho \; = \; {1 \over N} \sum_i
\langle \rho_i^+ + \rho_i^- \rangle
\end{equation}
which we assume to be homogeneous in space. The corresponding mean
field  Hamiltonian is given by:
\begin{widetext}
\begin{eqnarray}
H_{\rm MFA} & = & H_F +H_B - v \rho x + {\hbar \omega_0 \over 2}
\nonumber \\
H_F & = & (D-\mu)\sum_{i,\sigma}c^+_{i\sigma}c_{i\sigma}
-t\sum_{\langle i\neq j\rangle,\sigma}c^+_{i\sigma}c_{j\sigma} +
{v \rho \over 2}\sum_i [ c_{i\downarrow}c_{i\uparrow}
+c^+_{i\uparrow}c^+_{i\downarrow}] \nonumber \\
H_B & = & (\Delta_B-2\mu) \sum_i \left( \rho_i^z + \frac{1}{2}
\right) +v x \sum_i [\rho^+_i \; + \rho_i^-] - \hbar \omega_0
\alpha \sum_i \left( \rho_i^z + \frac{1}{2} \right) (a_i+a_i^{+})
+ \hbar \omega_0 \sum_i a^{+}_i a_i \; .
\end{eqnarray}
\end{widetext}

On the level of this approximation, the effect of the charge
fluctuations between bound pairs and pairs of itinerant electrons
is diminished, being taken into account in a global, spatially
independent, fashion. The itinerancy of the intrinsically
localized bound electron pairs, induced by the exchange coupling
\cite{Ranninger-95,Domanski-01b} is not contained in the present
approximation and therefore we cannot account for collective
soundwave-like excitations and the spatially correlated lattice
deformations which result from them. Nevertheless, the present
study does permit to investigate local excitations of the lattice,
controlled by the local fluctuations of the order parameters to
which they are coupled, and which can be related to a number of
experiments.

Considering itinerant bosons on a deformable lattice with weak
boson-phonon coupling\cite{Jackeli-01}, leads to a hybridization
of the Bogoliubov modes (coming from the repulsive interaction
between the bosons) and the Einstein-like phonon modes. At long
wavelengths this results in a modified sound velocity of the
Bogoliubov mode, with little spectral weight carried by the
phonons, and a moderately renormalized Einstein mode, principally
of phonon character. In such a context, what the present
mean-field study permits us to do, is to investigate the effect of
the superfluid state on such an Einstein-like mode. Due to the
strong coupling between the bound electron pairs and the phonons,
such an Einstein mode will undergo a substantial renormalization,
as it will be shown below.

The fermionic part $H_F$ of $H_{\rm MFA}$, being dynamically
decoupled from the bosonic part $H_B$ and hence from the phonons,
can be cast into the standard form $H_F = \sum_{\bf k}
\tilde\varepsilon_{\bf k}(\rho) \left( \alpha_k^+ \alpha_k \;+\;
\beta_k^+ \beta_k\right)$. The $\alpha$'s and $\beta$'s denote the
operators of Bogoliubov quasi-particles having the dispersion
$\tilde\varepsilon_{\bf k}(\rho) = \pm \sqrt{(\varepsilon_{\bf
k}-\mu)^2 + (v \rho)^2/4}$ which differs from the  bare electron
dispersion $\varepsilon_{\bf k}$ by showing a gap of size $v
\rho$. The corresponding eigenstates of $H_F$ are given by the
standard BCS expression
\begin{equation}
|\Psi^{\rm BCS} \rangle = \prod_k \left[ u_k + v_k c^+_{k
\uparrow}c^+_{-k \downarrow} \right]| 0 \rangle \; .
\end{equation}
The bosonic part $H_B$ of $H_{\rm MFA}$, showing intrinsic
coupling between the bosonic charge carriers and the phonons, is
diagonalized in terms of a set of states of the form
\begin{equation}
|\; l \;\}^{sc}_i = \sum_n \left[ u^{sc}_{l n}(i) +
v^{sc}_{ln}(i)\,\rho_i^+\right] | 0 )_i|n>_i
\end{equation}
having eigenvalues $E^{sc}_l(x)$. The eigenstates of the
mean-field Hamiltonian, Eq.\,(15), are thus given by
\begin{equation}
|\Psi^{\rm BCS} \rangle \otimes   \prod_i |\; l \;\}^{sc}_i \; .
\end{equation}

As in the atomic limit, the diagonalization of $H_B$ is carried
out in a truncated Hilbert space spanned by the eigenstates $|n>$
of the undeformed harmonic oscillator. We ultimately arrive at the
following set of self-consistent equations determining the
temperature dependence of the order parameters
\begin{eqnarray}
x &=& -{v \rho \over 4 N} \,\sum_{\bf k} \,{1 \over
\tilde\varepsilon_{\bf k}(\rho)}
\,\tanh{\beta \tilde\varepsilon_{\bf k}(\rho) \over 2} \nonumber \\
\rho &=& {1 \over Z}\, \sum_{ln} \, u^{sc}_{ln} \, v^{sc}_{ln} \,
\exp\,\left[-\beta E^{sc}_l(x)\right] \; ,
\end{eqnarray}
$Z=\sum_{l} \exp\,[-\beta E^{sc}_l(x)]$ denoting the partition
function corresponding to $H_B$. This set of equations must be
solved with the constraint that the total density of bosons and
fermions remains fixed, i.e. $n_{tot} = n_F + 2 n_B +
\frac{1}{4}\rho^2$, where
\begin{eqnarray}
n_F &=& {1 \over N}\sum_{{\bf k}\sigma} \langle c^+_{{\bf
k}\sigma}c_{{\bf k}\sigma} \rangle \;=\; {1 \over N}\sum_{{\bf k}}
\left( 1 - {\varepsilon_{\bf k} \over \tilde\varepsilon_{\bf
k}(\rho)}
\, \tanh{\beta \tilde\varepsilon_{\bf k}(\rho) \over 2}\right)\nonumber \\
n_B &=& {1 \over 2 } + {1 \over  N} \sum_i \langle \rho_i^z
\rangle \nonumber \\
&=& {1 \over 2 } + \frac{1}{2Z} \;\sum_{ln} \left[ (u^{sc}_{ln})^2
- (v^{sc}_{ln})^2 \right] \exp\,\left[-\beta E^{sc}_l(x)\right] \;
.
\end{eqnarray}

Let us now examine the modification of the composition of the
local coherent phonon states on the individual lattice sites
(discussed above on the basis of the atomic limit for the normal state)
when a macroscopic  phase-coherent state of the charge carriers emerges
and the bound elelctron-pairs condense into a superfluid state. The
probability distribution of the phonons in the superconducting phase
is then described by
\begin{equation}
P_{sc}(n) = {1 \over Z}\sum_l  \exp\,\left[-\beta
E^{sc}_l(x)\right] \left[ (u^{sc}_{ln})^2 + (v^{sc}
_{ln})^2\right] \; .
\end{equation}
The evaluation of $P_{sc}(n)$ in the superfluid phase follows
closely the calculation for $P_{at}(n)$ in the atomic limit. In
fact, the secular equation determining the coefficients
$u^{sc}_{ln}$ and $v^{sc}_{ln}$ is formally the same as in the
atomic limit, with the states $c^+_{\uparrow}c^+_{\downarrow}|0>
\otimes \,|0) \otimes |n>$ and $|0> \otimes \,\rho^+ |0) \otimes
|n>$ spanning the Hilbert space being now replaced by $ |0)
\otimes |n>$ and $\rho^+ |0) \otimes  |n>$, respectively. The only
modifications which then appear in this set of equations are ($i$)
a renormalized hybridization constant $\tilde v \equiv v x$,
instead of $v$ in the atomic limit, and ($ii$) an energy level for
the diagonal elements involving the states $| 0 ) \otimes | n >$
being equal to $\hbar \omega_0 n$, instead of $2(D-\mu)+\hbar
\omega_0 n $ for the diagonal elements involving the states
$c^+_{\uparrow}c^+_{\downarrow}|0> \otimes \,|0) \otimes |n>$.
This renormalization of $v$ is important since it reduces the bare
$v$ by roughly an order of magnitude (for the set of parameters
chosen throughout this paper),  expressing the fact that the
superfluidity leads to a considerable reduction of the on-site
correlation between the configurations with bound electron-pairs
present and with such pairs being absent. This has significant
consequences on the composition of the local phonon clouds
surrounding the individual sites and thus on their corresponding
PDFs.

Just below the superconducting $T_c$ ($\simeq$ 0.0067 in our
present case), $P_{sc}(n)$ is determined by an incoherent
superposition of essentially the two lowest eigenstates of the
mean-field Hamiltonian, Eq.\,(15), one giving rise to a broad peak
at some finite $n$ and  the other to a sharp maximum at $n=0$. At
low temperatures, however, the ground state alone is sufficient to
describe $P_{sc}(n)$. The contribution to $P_{sc}(n)$ leading to a
peak at some finite value of $n$ (associated with the coefficients
$v^{sc}_{0n}$) arises from the deformation induced by the presence
of the bound electron pairs, while the contribution to $P_{sc}(n)$
leading to a maximum for $n=0$ (associated with the coefficients
$u^{sc}_{0n}$) arises from the absence of such an induced
deformation for configurations where no bound pairs are present,
with the oscillators being essentially undeformed. The coherent
superposition of deformed and undeformed oscillator states in the
ground state is a direct consequence of the macroscopic phase
locking of the composite bosonic states in the superconducting
phase, resulting from the variational ground state of $H_B$ of the
form
\begin{eqnarray}
\prod_i \left ( e^{-i\phi_i/2} \widetilde X_i^-  + e^{i\phi_i/2}
\widetilde X_i^+ \rho_i^+ \right ) |0)_i \otimes|0>_i \; .
\end{eqnarray}
The operators $\widetilde X_i^{\pm}$ are defined as
\begin{eqnarray}
\widetilde X_i^{-}|0>_i&=&\sum_n u^{sc}_{0n}(i)|n>_i \\
\widetilde X_i^{+}|0>_i&=&\sum_n v^{sc}_{0n}(i)|n>_i
\end{eqnarray}
and differ significantly from the usual polaron shift operators
$X^{\pm}_i=e^{\mp \alpha(a_i -a_i^+)}$, characterizing the
deformations of a simply shifted oscillator state. {\it
Macroscopic} phase locking and homogeneity imply that the
coefficients $u^{sc}_{0 n}(i)$ and $v^{sc}_{0 n}(i)$ as well as
phases $\phi_i$ are the same for all sites. Eq.\,(22) is a direct
generalization of the superfluid ground state of a system of
hard-core bosons on a lattice, which is of the form given in
Eq.\,(13), and thus shows explicitly the locking together of the
charge $(\rho_i^+)$ and lattice ($\widetilde X_i^{\pm}$) degrees
of freedom on a local level.
\begin{figure}
\includegraphics[width=7cm]{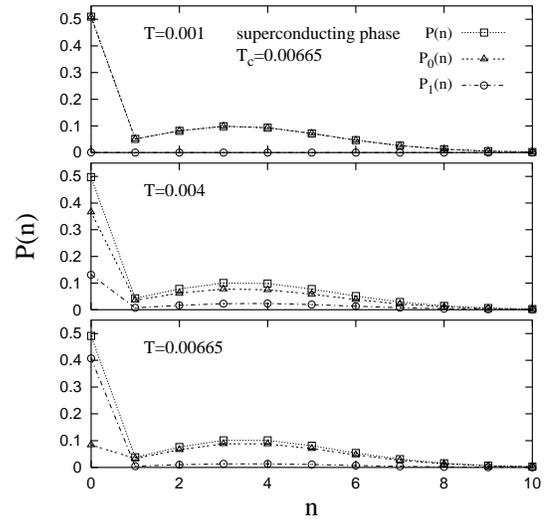}
\caption{Decomposition, for different temperatures, of $P(n)$
(squares) in the superconducting phase in terms of the
contributions $P_{0}(n)$ coming from the ground state (triangles)
and $P_{1}(n)$ coming from the first excited state (circles) of
$H_B$.} \label{PN2}
\end{figure}
We shall next address ourselves to the question of how such a
{\it macroscopic} coherent quantum state of the phonons can be tracked
experimentally and, in particular, what difference there is to be expected
with respect to the normal state features of the PDF, examined in the
previous section. The PDF in the superconducting phase is given by
\begin{equation}
g_{sc}(X) = {1 \over Z} \sum_l e^{-\beta E^{sc}_l(x)} \; _i^{sc}\{
l | \delta (X - X_i)  | l \}^{sc}_i \; .
\end{equation}
\begin{figure}
\includegraphics[width=7cm]{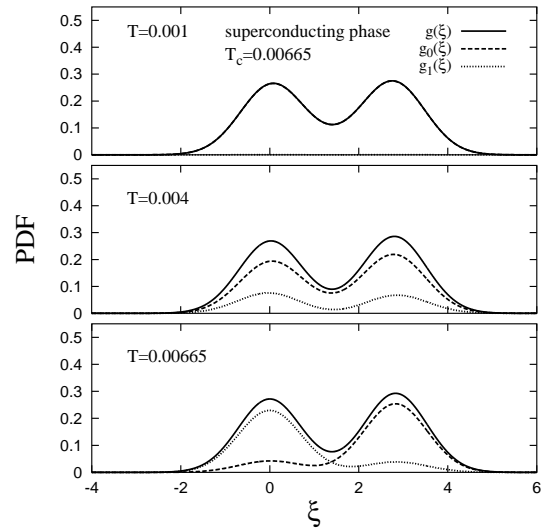}
\caption{Decomposition, for different temperatures, of $g(\xi)$ in
the superconducting phase (full lines) into the contributions
$g_0(\xi)$ coming from the ground state (dashed) and $g_1(\xi)$
coming from the first excited state (pointed) of $H_B$.}
\label{PDF2}
\end{figure}

In Fig.\,\ref{PDF0} we compare the PDF in the normal state and in
the superconducting phase for different temperatures. The features
examined above on the basis of the composition of $P_{sc}(n)$ are
perfectly reflected in the PDF. In the normal state one observes a
single peak at some finite $X$, indicative of well defined shifted
oscillator states. In the superconducting phase, on the contrary,
one observes a weakly temperature dependent double-peak structure,
which, as we can see from  Fig.\,\ref{PDF2}, arises essentially
from two contributions associated with the ground state and the
first excited state of $H_B$. At temperatures just below $T_c$
this double-peak structure results from an incoherent
superposition of these two states, corresponding to a situation
where localized bipolarons are randomly distributed over the
lattice. The statistical distribution of empty and doubly occupied
sites will thus give rise to that double-peak structure. In the
limit of low temperature, on the contrary, the PDF arises $\it
{exclusively}$ from the ground state, with a peak centered near
$X=0$ (being determined by the coefficients $u^{sc}_{0n}$) and a peak
centered around a finite value of $X$ (being determined by the
coefficients $v^{sc}_{0n}$). The appearence of a double-peak structure in
the PDF has long since been recognized as a signature of
delocalization of the charge carriers in polaronic
systems\cite{Shore-73}. The present study shows that, as the
superconducting state sets in and is strengthened upon lowering
the temperature, the {\it macroscopic} phase coherence of the
system not only involves the charge carriers but also the
deformations which surround them. This is only possible if at
$T=0$ these deformations are controlled  by a single quantum
state, i.e., the ground state $\prod_i|0\}_i^{sc}$. The
superconductivity-induced delocalization of the self-trapped
entities which characterize the normal state, is caused by the
locking together of the local lattice deformations in a coherent
superposition of deformed and essentially undeformed oscillator
states, resulting in the developement of a double-peak structure
in the PDF. At least in the scenario presented here, this result
does not bear out any undressings\cite{Hirsch-00} of the charge
carriers, stripping off partially the phonon clouds, as one enters
the superconducting phase. Such an effect would show up in a PDF
which, in the superconducting phase, would be given by essentially
a single-peak structure centered around $X=0$.

\section{Dynamical lattice properties}
The discussion in the previous Section demonstrated the
qualitative difference which exists between some basic lattice properties
in the normal and the superconducting phase, and largely being  controlled
by the induced electron pairing in the subsystem of the intrinsically
uncorrelated itinerant electrons. What results is a competition between
the local pairing, characterizing the normal-state phase, and the
{\bf k}-space pairing in the superconducting phase, expected to be
particularly relevant around the superconducting transition where
the spatial phase fluctuations play a major role and residual local
electron pairing\cite{Levin-98} persists in the superconducting phase.
The qualitative difference in the structure of the PDF in the two
phases leads us to expect squeezing effects of the local coherent
phonon quantum states which are stronger in the pseudogap phase
than in the superconducting phase. Associated with it, one should expect
larger fluctuations of the local deformations of the environment and of
the momenta of the ions representing it. These fluctuations are given by
\begin{equation}
\Delta X_i^2 = \langle X_i^2 \rangle - \langle X_i \rangle^2\; ,
\qquad \Delta P_i^2 = \langle P_i^2 \rangle
\end{equation}
with
\begin{eqnarray}
X_i & = & \sqrt{\frac{\hbar}{2 M \omega_0}}\;(a_i^++a_i) \nonumber
\\ P_i & = & i \sqrt{\frac{\hbar M\omega_0}{2 }}\;(a_i^+-a_i) \; .
\label{XP}
\end{eqnarray}

\begin{figure}
\includegraphics[width=7cm]{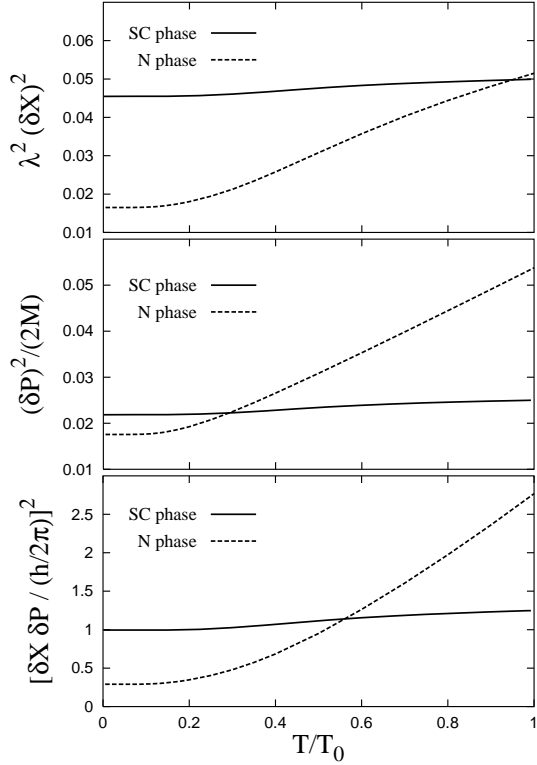}
\caption{Comparison of (a) the positional and (b) the momentum
fluctuations of the lattice in the normal (N) and in the
superconducting (SC) phase as a function of relative temperature
$T/T_0$ ($T_0=\{T_c,T^*\}$) and with $\lambda^2=(1/2) M\omega_0^2
\varepsilon_p$. (c) Degree of the squeezing of the coherent phonon
states as a function of $T/T_0$.} \label{DXDP}
\end{figure}

\begin{figure}
\includegraphics[width=7cm]{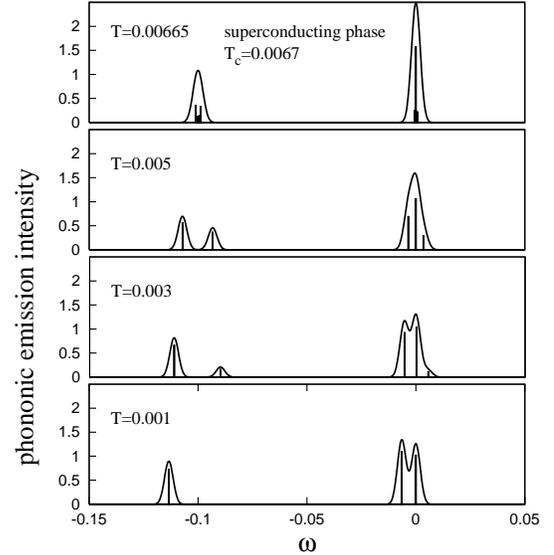}
\caption{Evolution of the phonon spectrum in emission experiments
as a function of temperature in the superconducting phase.}
\label{PH2}
\end{figure}

The expectation values in Eq.\,(\ref{XP}) have to be calculated
with respect to the eigenvectors of $H_{at}$ and $H_{B}$ for,
respectively, for the normal  and the superconducting phase. The
results for these fluctuations in the two phases are illustrated
in Fig.\,\ref{DXDP}. In particular, the product $\Delta X_i^2
\Delta P_i^2$ clearly shows that the coherent phonon states in the
normal state are considerably squeezed, i.e., this product is for
$T\rightarrow 0$ very close to the lower limit of the Heisenberg
uncertainty principle, equal to $\hbar^2/4$. In the
superconducting phase, on the contrary, we observe practically no
squeezing effect with decreasing temperature, the product $\Delta
X_i^2 \Delta P_i^2$ being always bigger than what would be
expected in the normal state in this temperature regime. A
possible experimental verification of these features comes from
neutron absorption measurements\cite{Mook-90} which measure the
kinetic energy $E_{kin}= \frac{1}{2M}\langle P_i^2 \rangle$ of the
local lattice vibrations. Indeed, the experimental results shown
that, as the temperature is reduced, $E_{kin}$ saturates in the
superconducting phase and is distinctly above the values expected
in the normal state at the same temperature. This fact is born out
in our study, as can be seen by inspection of Fig.\,\ref{DXDP}(b).
Finally, an issue which ought to be of crucial interest in
studying the lattice properties connected with the onset of the
superconducting order or of the electron pairing at $T^*$ in the
normal state, is the question of the changes in the phonon
spectral properties upon going from one phase to the other. Strong
changes in the frequency of specific local modes, induced by
strong electron-phonon coupling, have been well established by
photoinduced infrared absorption measurements\cite{Taliani-90} and
been related to strong local displacements of the ions. In this
context, the appearence of new additional modes occuring with
changes from the normal into the superconducting phase have also
been reported, although this is still a matter of
debate\cite{McQueeney-99,Pintschovius-99}. More established are
the strong intensity changes of certain Raman active modes and the
strong frequency renormalizations and lifetime broadening occuring
at such transitions\cite{Ruani-97,Misochko-99}. Such effects can
best be studied by energy loss spectra in inelastic neutron
scattering or Raman spectroscopy where phonons are emitted. The
cross-section for it is given by
\begin{eqnarray}
A_{\rm emi}(\omega) & = & \int dt\,e^{i\omega t} A_{\rm emi}(t)
\nonumber
\\ &=& {1\over Z} \sum_{l,m}e^{-\beta\varepsilon_l}
\,|<l|a|m>|^2\, \delta(\omega+\varepsilon_m-\varepsilon_l)
\label{PHemi}
\end{eqnarray}
with
\begin{equation}
A_{\rm emi}(t)  =  <a(0)\,a^{\dag}(t)>\,\theta(-t) \; .
\end{equation}
\begin{figure}
\includegraphics[width=7cm]{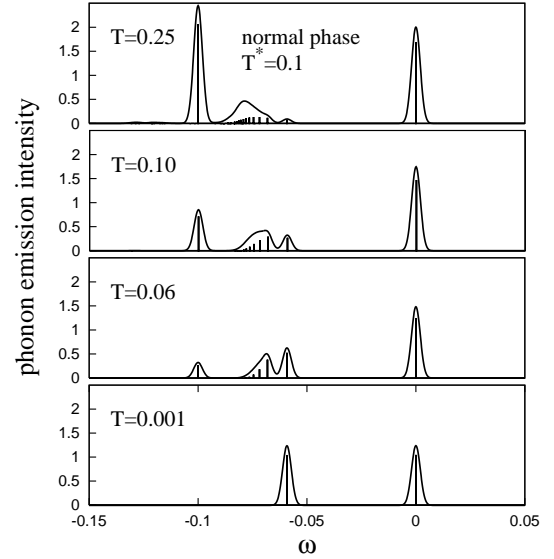}
\caption{Evolution of the phonon spectrum in emission experiments
as a function of temperature in the normal state.}
\label{PH1}
\end{figure}
\noindent Here $\varepsilon_l=\{E_l^{at},E_l^{sc}(x)\}$ and
$|l>=\{|m,l>^{at}, |l>^{sc}\}$ are the eigenvalues and
eigenvectors of the Hamiltonians $H_{at}$ and $H_B$, respectively.
We illustrate the results of such phonon emission spectra for the
two phases in Figs. \ref{PH1} and \ref{PH2}. Quite generally, one
observes a quasi-elastic contribution at frequency zero, which is
the standard signature of coherent phonon states. As the
temperature is lowered below either $T^*$ or $T_c$ and the
itinerant electrons get correlated by local pairing in the normal
state or by Cooper pairing in the superconducting phase, the local
phonon mode first splits into several modes and finally evolves
into a well defined mode which is softer in the pseudogap phase
and harder in the superconducting phase. These features of the
phonon spectral properties are expected to hold qualitatively
true. A  quantitative link-up between the normal phase and the
superconducting phase as the temperature is decreased, would
require a full description of the local boson-fermion exchange
correlations, incorporating the electron itinerancy in the normal
state and allowing for local electron pairing in the
superconducting phase. These are presently largely unresolved
fundamental problems in such system where amplitude and phase
order occur independently, as in the high-temperature
superconductors with its pseudogap phase above $T_c$ where local
electron pairing persists in an incoherent fashion.
\section{conclusions}
The present study is an attempt to examine the condensed state of
a system of composite particles such as charge carriers surrounded
by clouds of bosonic excitations of the neighboring environment to
which they are coupled and to see to what extent a macroscopic
coherent quantum state can be induced in the system of these
bosonic excitations. As a specific example we chose a system of
electrons, moderately strongly coupled to local lattice
deformations, which can be described in terms of resonant
local-pair states (localized bipolarons) inside a Fermi sea of
essentially uncorrelated itinerant electrons with which they
interact via a charge exchange term. We find noticeable
qualitative differences in the structure of the local coherent
quantum states of the phonons in the superconducting phase and in
the phase above it which corresponds to a normal state
characterized by a pseudogap in the DOS of the single-particle
spectrum. Several measurable quantities related to  static and
dynamic lattice properties were examined, such as the pair
distribution function, the fluctuations in the lattice positions
and the corresponding ion kinetic energies of the atoms making up
such a lattice and, finally, the phonon spectral properties given
by energy-loss spectroscopy (measurable by inelastic neutron
scattering or Raman spectroscopy). The difference between local
electron pairing in the normal state and {\bf k}-space pairing in
the superconducting phase has noticeable consequences on such
measurable quantities, resulting in ($i$) a single-peak structure
of the PDF in the normal state and a double-peak structure in the
superconducting phase, ($ii$) distinctively larger positional
fluctuations and kinetic energy of the atomic clusters (making up
the lattice) in the superconducting phase than in the normal state
and, ($iii$) associated with it, a more effective squeezing of the
phonon coherent quantum states in that latter. For the
superconducting phase, the present mean-field study is restricted
to local phase fluctuations of the superconducting state given by
a generalization of a phase-locked state of bosons on a lattice,
Eqs.\,(17,\,18,\,22), in which these bosons correspond to
composite particles in form of bipolarons [see Eq.\,(5)]. The
examination of soundwave-like collective excitations of such a
system is planned for some future work. There, the question arises
if the local dynamical deformations do get spatially correlated
upon entering the superconducting state, as possibly suggested by
ion channeling experiments\cite{Sharma-96}. A possible approach to
solve this problem could be to integrate out the fermionic sector
of the BFM Hamiltonian Eq.\,(1), resulting in a Gross-Pitaevskii
Lagrangian\cite{Geschkenbein-97} for the effective itinerant
bosonic charge carriers coupled to the phonons, followed by a
treatment similar to that adopted for bosons on a defromable
lattice\cite{Jackeli-01}.


\begin{thebibliography}{99}
\bibitem{Alexandrov-81} A.S. Alexandrov and J. Ranninger, Phys. Rev. B,
{\bf 23}, 1796 (1981).
\bibitem{Holstein-59} T. Holstein, Ann. Phys. (NY) {\bf 8}, 343 (1959).
\bibitem{deMello-97} E. de Mello and J. Ranninger, Phys. Rev. B {\bf 55},
14872 (1997).
\bibitem{Zhang-99} C. Zhang, E. Jeckelmann and S. White, Phys. Rev. B
{\bf 60}, 14092 (1999).
\bibitem{Ranninger-85} J. Ranninger and S. Robaszkiewicz, Physica B
{\bf 135}, 468 (1985).
\bibitem{Auerbach-01} E. Altman and A. Auerbach, cond-mat/0108087.
\bibitem{Kochetov-02} E. Kochetov and M. Mierzejewski, cond-mat/0204420.
\bibitem{Lee-94} R. Friedberg, T.D. Lee and H.K. Ren, Phys. Rev. B {\bf 50},
10190 (1994).
\bibitem{Yurovski-02} V.A. Yurovski and A. Ben-Reuven, cond-mat/0205267.
\bibitem{Fehske-97} H. Fehske, J. Loos and G. Wellein, Z. Phys. B
{\bf 104}, 619 (1997).
\bibitem{Stephan-97} W. Stephan, Phys. Rev. B, {\bf 54}, 8981 (1997).
\bibitem{Romero-99} A.H. Romero, D.W. Brown and K. Lindenberg,
Phys. Rev. B {\bf 59}, 13728 (1999).
\bibitem{Anderson-75} P.W. Anderson, Mater. Res. Bull. {\bf 8}, 153 (1976).
\bibitem{Schlenker-76} S. Lakkis, C. Schlenker, B.K. Chakraverty,
R. Buder and M. Marezio, Phys. Rev. B {\bf 14}, 1429 (1976).
\bibitem{Chakraverty-78} B.K. Chakraverty, J.M. Sienko and
J. Bonnerot, Phys. Rev. B {\bf 17}, 3781 (1978).
\bibitem{Salje-80} O.F. Schirmer and E. Salje, J. Phys. C
{\bf 13}, L1067 (1980).
\bibitem{Chernick-82} I.A. Chernick, S.N. Lykov and N.I. Grechko, Sov.
Phys. Solid State {\bf 24}, 1661 (1982).
\bibitem{Ranninger-98a} J. Ranninger and A. Romano, Phys. Rev. Lett.
{\bf 80}, 5643 (1998).
\bibitem{Domanski-01a} T. Domanski, Molecular Physics Reports
{\bf 34}, 39 (2001).
\bibitem{Egami-96} T. Egami and S.J.L. Billinge, in {\it Physical
properties of high temperature superconductors}, edited by D.
Ginsberg (World Scientific, Singapore, 1996) Vol. V, p. 265; J.
R\"ohler {\it et al.}, in {\it Workshop in high-$T_c$
superconductivity 1996: ten years after the discovery}, edited by
E. Kaldis {\it et al.} (Kluwer, Dordrecht, 1997) p. 469.
\bibitem{Ranninger-95} J. Ranninger, J.M. Robin and M. Eschrig, Phys. Rev.
Lett. {\bf 74}, 4027 (1995).
\bibitem{Ranninger-98b} J.M. Robin, A. Romano and J. Ranninger, Phys.
Rev. Lett. {\bf 81}, 2755 (1998).
\bibitem{Domanski-01b} T. Domanski and J. Ranninger, Phys. Rev. B
{\bf 63}, 134505 (2001).
\bibitem{Devillard-00} P. Devillard and J. Ranninger, Phys. Rev. Lett.
{\bf 84}, 5200 (2000).
\bibitem{Ranninger-99}  J. Ranninger and L. Tripodi, Solid State Commun.
{\bf 112}, 349 (1999).
\bibitem{Jackeli-01} G. Jackeli and J. Ranninger, Phys. Rev. B
{\bf 63}, 184521 (2001); {\it ibid.} {\bf 64}, 104513 (2001).
\bibitem{Shore-73} H.B. Shore and L.M. Sander, Phys. Rev. B {\bf 7}, 4537
(1973).
\bibitem{Hirsch-00} J. Hirsch, Phys. Rev. B {\bf 62}, 14487 (2000).
\bibitem{Levin-98} I. Kosztin, Q. Chen, B. Janko and K. Levin, Phys. Rev. B
{\bf 58}, R5936 (1998).
\bibitem{Taliani-90} C. Taliani {\it et al.}, in {\it Electronic properties of
High-$T_c$ Superconductors and related compounds},  edited by H.
Kuzmany, M Mehring and J. Fink (Springer-Verlag, Berlin, Series of
Solid State Science, 1990), p. 208.
\bibitem{McQueeney-99} R. J. McQueeney {\it et al.}, Phys. Rev. Lett. {\bf 82},
628 (1999).
\bibitem{Pintschovius-99} L. Pintschovius and M. Braden, Phys. Rev. B
{\bf 60}, R15039 (1999).
\bibitem{Ruani-97} G. Ruani and P. Ricci, Phys. Rev. B {\bf 55}, 93 (1997).
\bibitem{Misochko-99} O.V. Misochko {\it et al.}, Phys. Rev. B {\bf 59}, 11495
(1999).
\bibitem{Mook-90} H.A. Mook {\it et al.}, Phys. Rev. Lett. {\bf 65}, 2712 (1990).
\bibitem{Sharma-96} R.P. Sharma, T. Venkatesan, Z.H. Zhang, L.R. Liu,
R. Chu and W.K. Chu, Phys. Rev. Lett. {\bf 77}, 4624 (1996).
\bibitem{Geschkenbein-97} V.B. Geschkenbein, L.B. Ioffe and A.I.
Larkin, Phys. Rev. B {\bf 55}, 3173 (1997).

\end{thebibliography}
\end{document}